\documentclass{webofc}
\usepackage[varg]{txfonts}   
\usepackage{braket}
\usepackage{slashed}

\newcommand{\order}{\mathcal{O}}

\newcommand{\Dslash}{\slashed{D}}
\newcommand{\vslash}{\slashed{v}}

\newcommand{\kslash}{\slashed{k}}
\newcommand{\matrice}{\mathcal{M}}
\newcommand{\mupi}{\hat{\mu}_\pi^2}
\newcommand{\muG}{\hat{\mu}_G^2}
\newcommand{\rhoD}{\hat{\rho}_D^3}
\newcommand{\rhoLS}{\hat{\rho}_{LS}^3}

\begin{document}
\title{Inclusive rare $\Lambda_b$ decays to photon}
%
%

\author{\firstname{Francesco} \lastname{Loparco}\inst{1}\fnsep\thanks{\email{francesco.loparco1@ba.infn.it}}}

\institute{Istituto Nazionale di Fisica Nucleare, Sezione di Bari, Via Orabona 4, I-70126 Bari, Italy}

\abstract{
Inclusive $H_b \to X_s \, \gamma$ decays, with $H_b$ a beauty baryon, are treated exploiting an expansion up to the third order in the inverse of the $b$ quark mass $m_b$ and to LO in $\alpha_s$, keeping the dependence on the hadron spin. 
Double distribution $\frac{d^2\Gamma}{dy \, d \cos \theta_P}$ is computed for polarized baryons, with  $y = 2 \, E_\gamma / m_b$, $E_\gamma$ the photon energy and $\theta_P$ the angle between the  baryon spin vector and the photon momentum in the $H_b$ rest-frame.
Modifications to the photon polarization asymmetry can probe effects of physics beyond the Standard Model.
A systematic method to treat the singular terms in the photon energy spectrum obtained by the OPE is proposed.
}

\maketitle

\vspace{-0.7cm}

\section{Introduction}


Processes induced at the quark level by the $b \to s \, \gamma$ transition occur at loop level in the Standard Model (SM) and are therefore sensitive to physics beyond it \cite{Shifman:1976de,Bertolini:1986th,Grinstein:1987vj}.
Upon integration of the heavy quanta, an effective Hamiltonian is obtained in terms of local operators and Wilson coefficients \cite{Cella:1994px,Misiak:2015xwa}.
Physics beyond the Standard Model can induce new operators and/or modify the Wilson coefficients \cite{Grinstein:1987pu,Hou:1987kf,Gabbiani:1996hi,Everett:2001yy,Borzumati:2003rr,Buras:2003mk,Blanke:2006sb,Misiak:2017bgg}.

The radiative $b \to s$ transition has been analyzed in theory \cite{Buras:2020xsm}.
As for experiment, after the first observation  of the $B \to K^*(892) \, \gamma$ mode \cite{CLEO:1993nic} other exclusive processes have been observed \cite{HFLAV:2022pwe,ParticleDataGroup:2022pth}. 
For baryons,   the rate and the photon polarization  of $\Lambda_b \to \Lambda \, \gamma$ have been measured \cite{LHCb:2019wwi,LHCb:2021byf}, and  an upper bound has been put to ${\cal B}(\Xi_b^- \to \Xi^- \, \gamma)$ \cite{LHCb:2021hfz}.

Considering the inclusive mode $H_b \to X_s \, \gamma$, several analyses have been carried out for $B$ mesons.
In this paper we focus on the case of beauty baryons, in particular $\Lambda_b$ \cite{Colangelo:2023xnu}.
Invoking quark-hadron duality, it is possible to exploit the Operator Product Expansion (OPE) and Heavy Quark Effective Theory (HQET) \cite{Neubert:1993mb} to express inclusive decay widths as an expansion in $1 / m_b$.
Input quantities are the hadronic matrix elements of  local operators.

Several issues in inclusive modes induced by the $b \to s \, \gamma$ transition need to be considered.
The actual expansion parameter is the inverse of the energy released in the  process.
In some regions of the phase space it is not small.
Signals about the reliability of the method show up as  singularities in differential distributions.
Singular terms appear at higher orders in the Heavy Quark Expansion (HQE) in form of the delta distribution and its derivatives.
The origin of such terms can be related to the Fermi motion of the heavy quark in the decaying hadron, and can be accounted for introducing a shape function which encodes information on the distribution of the $b$
quark  residual momentum in   the hadron \cite{Neubert:1993ch,Neubert:1993um,Bigi:1993ex}.
In case of $B$,  measurements have been exploited to constrain its moments \cite{Bernlochner:2020jlt}.

Another source of uncertainty in inclusive $b \to s \gamma$ processes are the resolved photon contributions  \cite{Benzke:2010js}, that appear  at ${\cal O}(m_b^{-1})$ \cite{Donoghue:1995na,Lee:2006wn}. 

The photon energy spectrum can be measured above an energy threshold.
For $\bar B$ the HFLAV Collaboration provides ${\cal B}({\bar B} \to X_s \, \gamma) = ( 3.49 \pm 0.19 ) \,\times \, 10^{-4}$ for $E_\gamma > 1.6 \, \text{GeV}$
to be compared to the SM result: ${\cal B}({\bar B} \to X_s \, \gamma) = ( 3.36 \pm 0.23) \, \times \, 10^{-4}$ for the same  threshold \cite{Misiak:2015xwa}.

Baryon modes, such as $\Lambda_b \to X_s \, \gamma$ provide the further possibility to investigate observables sensitive to the spin of the decaying hadron. 
Heavy baryons with a b-quark produced in $Z^0$ and top-quark decays are expected to have a sizable polarization, as observed at LEP \cite{BUSKULIC1996437,Abbiendi:1998uz,2000205}.
The application of HQE to baryons requires new information, namely the operator matrix elements for specified hadron spin.
These have been determined in \cite{Colangelo:2020vhu}.
Here we also investigate a method to treat the singular terms in the inclusive photon spectrum to reconstruct the  $\Lambda_b$ shape function, a method which can be systematically applied when higher order terms in the $1 / m_b$ expansion are computed.

The plan of the paper is the following.
Section \ref{sec:Hamil} describes the $b \to s \, \gamma$ low-energy Hamiltonian both in SM and in extensions. 
In section \ref{OPE} we review the  application of the  HQE to the inclusive $H_b \to X_s \, \gamma$ process with $H_b$ a baryon, keeping the dependence on the baryon spin. 
The correlation between the photon and $\Lambda_b$ polarizations is studied.
A treatment of the singular terms is discussed in section \ref{sf}.
The last section summarizes.

\vspace{-0.2cm}

\section{Generalized $b \to s \, \gamma$ effective Hamiltonian}
\label{sec:Hamil}

The low-energy Hamiltonian governing
$b \to s \, \gamma$ transition can be written as
\begin{align}
\label{hamil}
H_{\rm eff}^{b \to s \gamma} = - 4 \,\frac{G_F}{\sqrt{2}} \, V_{tb} \, V_{ts}^* \, \sum_{i} \, \left[ C_i(\mu) \, O_i + C_i^{\prime}(\mu) \, O_i^{\prime} \right] \;,
\end{align}
$G_F$ is the Fermi constant, $V_{tb(s)}$ the CKM matrix elements.\footnote{Doubly Cabibbo suppressed terms proportional to $V_{ub} \, V_{us}^*$ have been neglected in \eqref{hamil}.}
Eq. \eqref{hamil} comprises the magnetic penguin operators $O_7 = \frac{e}{16 \, \pi^2} \, [ {\bar s} \, \sigma^{\mu \nu} \, ( m_s \, P_L + m_b \, P_R )\,b] \, F_{\mu \nu}$ and $O_8 = \frac{g_s}{16 \, \pi^2} \, \Big[ {\bar s}_{ \alpha} \, \sigma^{\mu \nu} \, \Big( \frac{\lambda^a}{2} \Big)_{\alpha \beta} \, ( m_s \, P_L + m_b \, P_R ) \, b_{\beta} \Big] \, G^a_{\mu \nu}$, with $P_{R,L} = \frac{1 \pm \gamma_5}{2}$ helicity projectors,  $\alpha,\beta$ colour indices, $\lambda^a$ the Gell-Mann matrices.
$F_{\mu \nu}$ and $G^a_{\mu \nu}$  are the electromagnetic and gluonic field strengths,  $e$ and $g_s$ the
electromagnetic and strong coupling constants, $m_b$ and $m_s$  the $b$ and $s$ quark mass. 
Current-current operators $O_1 = ( {\bar s}_\alpha \, \gamma^\mu \, P_L \, c_\beta ) \, ( {\bar c}_\beta \, \gamma_\mu \, P_L \, b_\alpha )$ and $O_2 = ( {\bar s} \, \gamma^\mu \, P_L \, c ) \, ( {\bar c} \, \gamma_\mu \, P_L \, b)$
and the  QCD penguin operators $O_3 = ( {\bar s} \, \gamma^\mu \, P_L \, b) \, \sum_q \, ( {\bar q} \, \gamma^\mu \, P_L \, q)$, $O_4 = ( {\bar s}_\alpha \, \gamma^\mu \, P_L \, b_\beta ) \, \sum_q \, ( {\bar q}_\beta \, \gamma^\mu \, P_L \, q_\alpha)$, $O_5 = ( {\bar s} \, \gamma^\mu \, P_L \, b) \, \sum_q \, ( {\bar q} \, \gamma^\mu \, P_R \, q )$ and $O_6 = ( {\bar s}_\alpha \, \gamma^\mu \, P_L \, b_\beta ) \, \sum_q \, ( {\bar q}_\beta \, \gamma^\mu \, P_R \, q_\alpha )$, where the sum runs over $q = \{ u, d, s, c, b \}$, are also comprised.
The remaining operators,  absent in SM,  are analogous to the QCD penguins but have a scalar or tensor structure \cite{Borzumati:1999qt}.
The primed operators  have opposite chirality with respect to the unprimed ones.

In SM the process $b \to s \, \gamma$ is described by photon penguin diagrams, with the photon coupled either to the intermediate fermion or to the $W^\pm$,  giving rise to the magnetic operator $O_7$, the only operator contributing at lowest order in QCD.
The renormalization group evolution to the scale $\mu_b \simeq {\cal O}(m_b)$ also involves $O_8$ and $O_{1,\dots 6}$.
Large logarithms  producing a strong enhancement of the rate are generated by the mixing of these operators into $O_7$.
The anomalous dimension matrix governing the mixing is regularization scheme dependent.
Such dependence is taken into account by defining an effective coefficient $C_7^{\rm eff}(\mu_b)$ which includes contributions of $O_{1,\dots 6}$ \cite{Buras:1993xp}.
Therefore, $O_7$ represents the dominant contribution to $b \to s \, \gamma$, with the SM Wilson coefficients known at NNLO in QCD \cite{Misiak:2018cec,Buras:2020xsm}.

In this study we describe a calculation at LO in $\alpha_s$ so that only the operators $O_7$ and $O_7'$ have been considered.
Hence, the effective Hamiltonian at the scale $\mu_b$ can be written as
\begin{align}
\label{hnew}
H_{\rm eff}^{b \to s \, \gamma} = - 4 \, \frac{G_F}{\sqrt{2}} \, V_{tb} \, V_{ts}^* \, \big\{ C_7^{\rm eff} \,O_7 + C_7^{\prime \rm eff} \, O_7^\prime \big\} = -  4 \, \frac{G_F}{\sqrt{2}} \, \lambda_t \, \frac{e}{16 \, \pi^2} \,\sum_{i = 7,7^\prime} \, C_i^{\rm eff} \, J^{i}_{\mu \nu} \, F^{\mu \nu} \;,
\end{align}
where $\lambda_t = V_{tb} \, V_{ts}^*$, $J^{i}_{\mu \nu} = [ {\bar s} \, \sigma_{\mu \nu} \, ( m_s \, ( 1 - P_i ) + m_b \, P_i ) \, b ]$ and $P_7 = P_R$ and $P_{7'} = P_L$.

\vspace{-0.4cm}

\section{Inclusive decay width}
\label{OPE}

\begin{figure}[h]
\centering
\sidecaption
\includegraphics[width=4.7cm,clip]{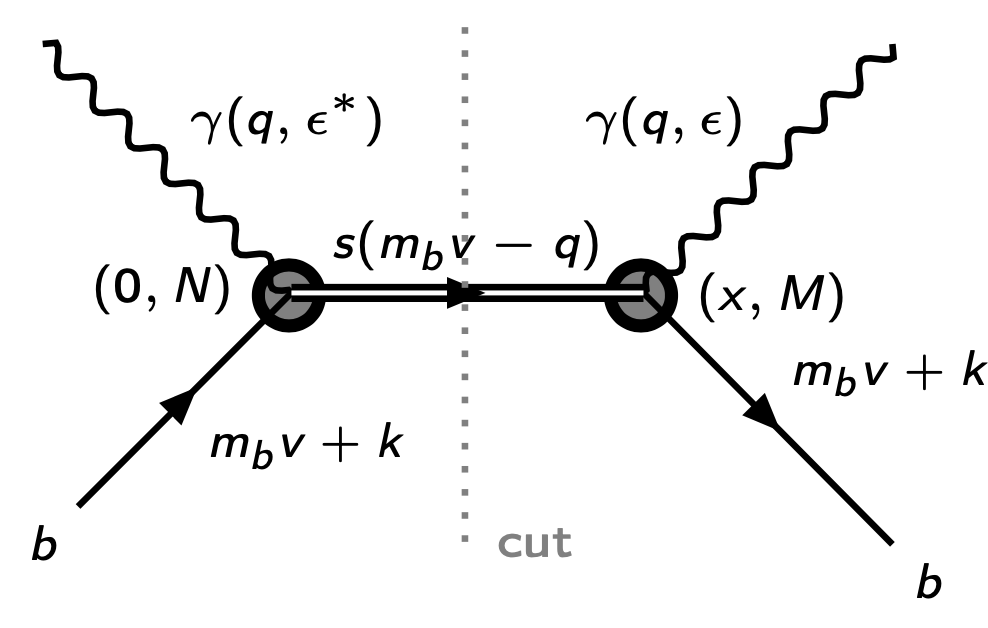}
\caption{Discontinuity of the forward amplitude across the cut of the radiative process.}
\label{discontinuity}
\end{figure}

The differential inclusive decay width can be written as
\begin{align}
\label{dgamma}
d\Gamma = [dq] \, \frac{G_F^2 \, |\lambda_t|^2}{8 \, m_{H_b}} \, \frac{\alpha}{\pi^2} \,\sum_{i,j=7,7^\prime} C_i^{\rm eff * } \, C_j^{\rm eff} \,W^{ij}_{MN} \, {\cal F}^{MN} \;,
\end{align}
with $[dq] = \frac{d^3 q}{(2 \, \pi)^3 \, 2 \, q^0}$ and ${\cal F}^{MN} = - 4 \, q^\nu \, q^{\nu^\prime} \, g^{\mu \mu^\prime}$.
By the optical theorem,  the hadronic tensor $W^{ij}_{MN}$ is  related to the discontinuity of the forward   scattering amplitude depicted in Fig. \ref{discontinuity}
\begin{align}
\label{Tij-gen}
T^{ij}_{MN} = i\,\int d^4 x \, e^{-i\,q \cdot x} \, \langle H_b(p,s) | T [ J^{i\dagger}_M (x) \,J^{j}_N (0) ] | H_b(p,s) \rangle \;,
\end{align}
across the cut corresponding to the process $H_b(p,s) \to X_s(p_X) \, \gamma (q,\epsilon)$: $W^{ij}_{MN} = \frac{1}{\pi} \, {\rm Im} \, \big[ T^{ij}_{MN} \big]$.
\vspace{-0.3cm}
\vspace{-0.5cm}
The range of the invariant mass $p_X^2$  of the states produced in $B$ and $\Lambda_b$ decays (with $p_X = p - q$) is $p_X^2 \in [ m^2_{K^*}, m_B^2 ]$ and $p_X^2 \in [ m_\Lambda^2, m_{\Lambda_b}^2 ]$, respectively. 
For $m_b \to \infty$, $p_X^2$ is almost always large enough to exploit the short-distance limit $x \to 0$ in Eq. \eqref{Tij-gen}, thus allowing a computation of $T^{ij}$ and $W^{ij}$ by an OPE  with expansion parameter $1 / m_b$ \cite{Chay:1990da,Bigi:1993fe}, which
can be constructed by expressing the hadron momentum $p = m_H \, v$, where the $v$ the four-velocity, in terms of $m_b$ and of a residual momentum $k$: $p = m_b \, v+k$.
The QCD $b$ quark field is rescaled $b(x) = e^{-i\,m_b v \cdot x} \, b_v(x)$,
with $b_v(x)$ and satisfies the equation of motion $b_v(x) = \big( P_+ + \frac{i \, \Dslash}{2 \, m_b} \big) \, b_v(x)$
with $P_+ = \frac{1 + \vslash}{2}$.
In terms of $b_v(x)$ Eq. \eqref{Tij-gen} becomes:
\begin{align}
T^{ij}_{MN} = i\,\int d^4 x \, e^{i\,( m_b v - q ) \cdot x} \, \langle H_b(v,s) | T [ {\hat J}^{i\dagger}_M (x) \, \hat J^{j}_N (0) ] | H_b(v,s) \rangle \;.
\end{align}
${\hat J}^{i}$ contains the field $b_v$. 
The HQE  is obtained from 
\begin{align}
T^{ij}_{MN} = \langle H_b(v,s) | {\bar b}_v(0) \, {\bar \Gamma}_M^{i} \, \frac{1}{m_b \, \vslash + \kslash - \slashed{q} - m_s} \,\Gamma_N^{j} \, b_v(0) |H_b(v,s) \rangle \;,
\end{align}
with $\bar \Gamma^{i}_M = \gamma^0 \, \Gamma^{i\dagger}_M \, \gamma^0$ and $\Gamma_M^{7} =\sigma_{\mu \nu} \,( m_b \, P_R + m_s \, P_L )$, $\Gamma_M^{7^\prime} = \sigma_{\mu \nu} \, ( m_b \, P_L + m_s \, P_R)$.
Replacing $k \to i \, D$, with $D$ the QCD covariant derivative, and considering $|k| \sim \mathcal{O}(\Lambda_\text{QCD})$ we have
\begin{align}
T^{ij}_{MN} = \sum_{n=0}^{+ \infty} \, \braket{H_b(v, s) | \bar{b}_v(0) \, \overline{\Gamma}_{M}^i \, (\slashed{p}_s + m_s) \, [ i \, \Dslash \, (\slashed{p}_s + m_s) ]^n \, \Gamma_{N}^j \, b_v(0) | H_b(v, s)} \, \frac{(-1)^n}{\Delta_0^{n+1}} \;,
\end{align}
where $p_s$ is the $s$ quark momentum and $\Delta_0 = p_s^2 - m_s^2$.
Using the trace formalism \cite{Dassinger:2006md}, we can write the n-th term in the series as
\vspace{-0.1cm}
\begin{align}
\label{trace_formalism}
& \braket{H_b(v, s) | \bar{b}_v(0) \, \overline{\Gamma}_M^i \, (\slashed{p}_s + m_s) \, \underbrace{i \, \Dslash \, (\slashed{p}_s + m_s) \dots i \, \Dslash \, (\slashed{p}_s + m_s)}_{\text{n times}} \, \Gamma_N^j \, b_v(0) | H_b(v, s)} = \\
				& = \bigg[ \overline{\Gamma}_M^i \, ( \slashed{p}_s + m_s ) \, \prod_{k=1}^n \, \left[ \gamma_{\mu_k} \, (\slashed{p}_s + m_s) \right] \, \Gamma_N^j \bigg]_{ab} \, \underbrace{\braket{H_b(v, s) | \bar{b}_v(0) \, i D^{\mu_1} \, \dots \, i D^{\mu_n} \, b_v(0) | H_b(v, s)}_{ba}}_{(\matrice^{\mu_1 \dots \mu_n})_{ba}} \;, \notag
\end{align}
with $a$ and $b$ Dirac indices.
For any order in the expansion, all the matrix elements $(\matrice^{\mu_1 \dots \mu_n})_{ba}$ can be written in terms of nonperturbative parameters
\vspace{-0.2cm}
\begin{equation}
\order(m_b^{- n}) \dots
\begin{cases}
\order(m_b^{- 3})
\begin{cases}
\order(m_b^{- 2})
\begin{cases}
- 2 \, M_H \, \mupi = \braket{H_b|\bar{b}_v \, i D^\mu \, i D_\mu \, b_v|H_b} \\
2 \, M_H \, \muG = \braket{H_b|\bar{b}_v \, (- i \sigma_{\mu\nu}) \, i D^\mu \, i D^\nu \, b_v|H_b}
\end{cases} \\
2 \, M_H \, \rhoD = \braket{H_b|\bar{b}_v \, i D^\mu \, (i v \cdot D) \, i D_\mu \, b_v|H_b} \\
2 \, M_H \, \rhoLS = \braket{H_b|\bar{b}_v \, (- i \sigma_{\mu\nu}) \, i D^\mu \, (i v \cdot D) \, i D^\nu \, b_v|H_b}
\end{cases} \\
\dots
\end{cases} \;.
\end{equation}
$\matrice^{\mu_1 \dots \mu_n}$ have been derived at $\order(m_b^{- 3})$ for a polarized baryon in \cite{Colangelo:2020vhu}, extending the previous results \cite{Mannel:2017jfk, Kamali:2018bdp, Grossman:1994ax, Manohar:1993qn, Balk:1997fg}.

The double distribution, with respect to $y$ and $\cos \theta_P$, using the methods proposed in \cite{Dassinger:2006md, Mannel:2010wj}, comprises two terms
\vspace{-0.2cm}
\begin{align}
\label{double}
\frac{d^2 \Gamma}{dy \, d \cos \theta_P} & = {\tilde \Gamma}_1 + \cos \theta_P \,{\tilde \Gamma}_2 \;.
\end{align}
Integrating \eqref{double} in $\cos \theta_P$, ($\theta_P$ being the angle between the hadron spin $s$ and the photon momentum $q$), one has
${\tilde \Gamma}_1 = \frac{1}{2} \, \frac{d \Gamma}{dy}$ and
the photon energy spectrum
\vspace{-0.2cm}
\begin{align}
\label{photonspectrum}
\frac{1}{\Gamma_0} \, \frac{d \Gamma}{d y} & = \bigg[ 1 - \frac{\mupi}{2 \, m_b^2} - \frac{\muG}{2 \, m_b^2} \, \frac{3 + 5 \, z}{1 - z} - \frac{10 \, \rhoD}{3 \, m_b^3} \, \frac{1 + z}{1 - z} \bigg] \, \delta ( 1 - z - y )  \notag \\
& \hspace{0.5cm} + \bigg[ \frac{\mupi}{2 \, m_b^2} \, ( 1 - z ) - \frac{\muG}{6 \, m_b^2} \, ( 3 + 5 \, z ) - \frac{4 \, \rhoD}{3 \, m_b^3} \, ( 1 + 2 \, z ) + \frac{2 \, \rhoLS}{3 \, m_b^3} \, ( 1 + z ) \bigg] \, \delta' ( 1 - z - y ) \notag \\
& \hspace{0.5cm} + \bigg[ \frac{\mupi}{6 \, m_b^2} \, ( 1 - z )^2 - \frac{\rhoD}{3 \, m_b^3} \, ( 1 - z ) \, ( 1 + 2 \, z ) + \frac{\rhoLS}{6 \, m_b^3} \, ( 1 - z^2 ) \bigg] \, \delta'' ( 1 - z - y ) \notag \\
& \hspace{0.5cm} - \frac{\rhoD}{18 \, m_b^3} \, ( 1 - z )^2 \, ( 1 + z ) \, \delta''' ( 1 - z - y ) \;,
\end{align}
where $z = \frac{m_s^2}{m_b^2}$, $\Gamma_0 = \frac{\alpha \, G_F^2 \, | \lambda_t |^2}{32 \, \pi^4} \, m_b^5 \, ( 1 - z )^3 \, \left[ | C_+^{\rm eff} |^2 + | C_+^{\prime \, \rm eff} |^2 \right]$, $C_+^{\rm eff} = C_7^{\rm eff} + \sqrt{z} \, C_7^{\prime \,\rm eff}$ and $C_+^{\prime \, \rm eff} = \sqrt{z} \, C_7^{\rm eff} + C_7^{\prime \,\rm eff}$.
${\tilde \Gamma}_2  $ is given by:
\vspace{-0.2cm}
\begin{align}
- \frac{2}{\Gamma_0} \, \frac{| C_+^{\rm eff} |^2 + | C_+^{\prime \, \rm eff} |^2}{| C_+^{\rm eff} |^2 - | C_+^{\prime \, \rm eff} |^2} \, \tilde{\Gamma}_2 & = \bigg[ 1 - \frac{13 \, \mupi}{12 \, m_b^2} - \frac{3 \, \muG}{4 \, m_b^2} \, \frac{5 + 3 \, z}{1 - z} - \frac{\rhoD}{6 \, m_b^3} \, \frac{31 + 9 \, z}{1 - z} \bigg] \, \delta ( 1 - z - y )  \notag \\
& \hspace{0.5cm} + \bigg[ \frac{\mupi}{2 \, m_b^2} \, ( 1 - z ) - \frac{\muG}{2 \, m_b^2} \, ( 3 + z ) - \frac{2 \, \rhoD}{m_b^3} \, ( 1 + z ) \bigg] \, \delta' ( 1 - z - y ) \notag   \\
& \hspace{0.5cm} + \bigg[ \frac{\mupi}{6 \, m_b^2} \, ( 1 - z )^2 - \frac{\rhoD}{3 \, m_b^3} \, ( 1 - z ) \, ( 1 + 2 \, z ) \bigg] \, \delta'' ( 1 - z - y )   \notag \\
& \hspace{0.5cm} - \frac{\rhoD}{18 \, m_b^3} \, ( 1 - z )^2 \, ( 1 + z ) \, \delta''' ( 1 - z - y ) \;.
\end{align}
The  angular differential distribution also comprises two terms:
\begin{align}
\label{angular}
\frac{d \Gamma (H_b \to X_s \, \gamma)}{d \cos \theta_P}  = A + B \, \cos \theta_P \;,
\end{align}
with $A = \frac{1}{2} \, \Gamma (H_b \to X_s \, \gamma)$ and $B = - \frac{\Gamma_0}{2} \, \frac{| C_+^{\rm eff} |^2 - | C_+^{\prime \, \rm eff} |^2}{| C_+^{\rm eff} |^2 + | C_+^{\prime \, \rm eff} |^2} \, \big[ 1 - \frac{13 \, \mupi}{12 \, m_b^2} - \frac{3 \, \muG}{4 \, m_b^2} \, \frac{5 + 3 \, z}{1 - z} - \frac{\rhoD}{6 \, m_b^3} \, \frac{31 + 9 \, z}{1 - z} \big]$.\\
Integrating in $\cos \theta_P$, the  inclusive  $H_b \to X_s \, \gamma$ width  is given by
\begin{align}
\label{totalG}
\Gamma (H_b \to X_s \, \gamma)  = \Gamma_0 \, \bigg[ 1 - \frac{\mupi}{2 \, m_b^2} - \frac{\muG}{2 \, m_b^2} \, \frac{3 + 5 \, z}{1 - z} - \frac{10 \, \rhoD}{3 \, m_b^3} \, \frac{1 + z}{1 - z} \bigg] \;.
\end{align}
For $C_7^\prime \to 0$ the SM result is recovered. 

\begin{figure}[h]
\begin{center}
\sidecaption
\includegraphics[width = 0.45\textwidth]{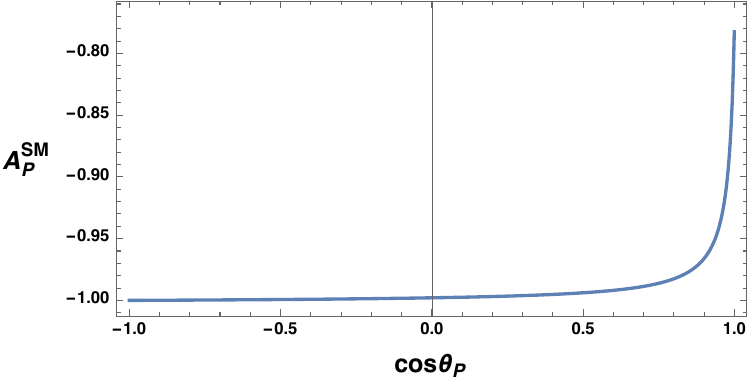}
\includegraphics[width = 0.52\textwidth]{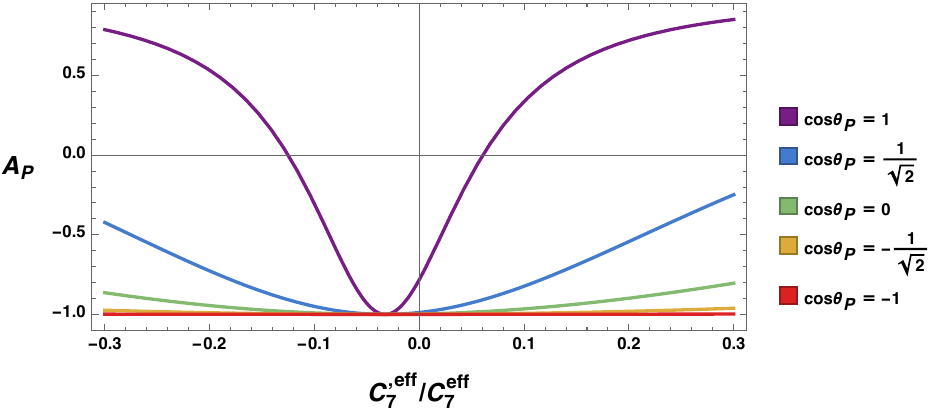}
\caption{\small  Photon polarization asymmetry $A_P(\cos \theta_P)$ versus $\cos \theta_P$ in SM (left) and  photon polarization asymmetry $A_P(\cos \theta_P)$ for several values of $\cos \theta_P $, varying $C^{\prime \, {\rm eff}}_7 / C_7^{\rm eff}$ (right).}
\label{APSM}
\end{center}
\end{figure}

Another interesting observable is the photon polarization asymmetry $A_P(\cos \theta_P)$ in $b \to s \, \gamma$ transition that quantifies the difference between left and right handed photon produced (the definition can be found in \cite{Colangelo:2023xnu}).
In SM the photon polarization asymmetry is $A_P(\cos \theta_P) \simeq - 1$ for almost  all  $\cos \theta_P$,  it  increases only for $\cos \theta_P \to 1$,  see Fig. \ref{APSM} (left panel).
Physics beyond SM can produce a deviation of the asymmetry from the SM value with the largest effect for $\cos \theta_P \simeq 1$.
Considering both $C_7$ and $C_7'$ real and varying $C^{\prime \,\rm eff}_7 / C_7^{\rm eff}  \in [-0.3,0.3]$ we obtain the asymmetry in Fig. \ref{APSM} (right panel) versus $C^{\prime \,\rm eff}_7 / C_7^{\rm eff} $ for selected values of $\cos \theta_P$.

\vspace{-1.4cm}

\section{Treatment of the  singular terms}
\label{sf}

The spectrum obtained by the short-distance OPE  does not account for possible smearing close to the end point region.
In this region, the Fermi motion of the $b$ quark and its soft interactions with the light degrees of freedom in the hadron  are relevant.
It is possible to resum the singular terms defining the spectral function
\cite{Bigi:1993fe,Bigi:1993ex,Neubert:1993ch,Neubert:1993um,Mannel:1994pm}
\begin{align}
\label{SF}
S_s(y) = \sum_{n = 0}^{\infty} \, \frac{M_n}{n!} \, \delta^{(n)} ( 1 - z - y ) \;,
\end{align}
and introducing the quantity $f(k_+)$ as\footnote{Perturbative corrections to the shape function and  its moments are discussed in \cite{Bauer:2003pi,Bosch:2004th}.}
\begin{align}
\label{intfk}
S_s(y) = \int d k_+ \, \delta \left( 1 - y - z + \frac{k_+}{m_b} \right) \, [ f(k_+) + {\cal O}(m_b^{-1}) ] \;.  
\end{align}

The photon energy spectrum  is obtained by the convolution \cite{Neubert:1993um}
\begin{align}
\label{convolution}
\frac{d \Gamma}{dy} = \int dk_+ \, f(k_+) \, \frac{d \Gamma}{dy}^* \;.
\end{align}
In $\frac{d \Gamma}{dy}^*$ the b quark mass $m_b$ is replaced by $m_b^* = m_b + k_+$, an exact  substitution at tree level.
For $k_+ \in [ - m_b, m_{H_b} - m_b]$, replacing $m_b \to m_b^*$ in  the variable $y$,  one finds ($m_s = 0$ to simplify the discussion): $y \to \frac{2 \, E_\gamma}{( m_b + k_+ )}$.  
Hence,  for $k_+^{\rm max} = m_{H_b} - m_b$ the maximum photon energy is $E_\gamma = \frac{m_{H_b}}{2}$ (physical end-point). 
The  shape function provides an  interpretation of the singular terms in the photon energy spectrum obtained  in the previous section. 
Writing Eq. \eqref{photonspectrum} as
\begin{align}
\label{dGdy}
\frac{1}{\Gamma} \, \frac{d \Gamma}{d y} = \sum_{n = 0}^3 \, \frac{M_n}{n!} \, \delta^{(n)} ( 1 - z - y ) \;,
\end{align}
with
$\Gamma$ in Eq. \eqref{totalG} and the moments
\begin{align}
\label{momspectrumSF}
M_0 & = 1 \;,
\hspace{0.8cm}
M_1 = \frac{\mupi}{2 \, m_b^2} \, ( 1 - z ) - \frac{\muG}{6 \, m_b^2} \, ( 3 + 5 \, z ) - \frac{4 \, \rhoD}{3 \, m_b^3} \, ( 1 + 2 \, z ) + \frac{2 \, \rhoLS}{3 \, m_b^3} \, ( 1 + z ) \;, \\
M_2 & = \frac{\mupi}{3 \, m_b^2} \, ( 1 - z )^2 - \frac{2 \, \rhoD}{3 \, m_b^3} \, ( 1 - z ) \, ( 1 + 2 \, z ) + \frac{\rhoLS}{3 \, m_b^3} \, ( 1 - z^2 ) \;,
\hspace{0.2cm}
M_3 = - \frac{\rhoD}{3 \, m_b^3} \, ( 1 - z )^2 \, ( 1 + z ) \;. \notag
\end{align}
Each moment $M_n$ in Eqs. \eqref{momspectrumSF}, has an expansion $M_n = \sum_{k=n}^\infty \, \frac{M_{n,k}}{m_b^k}$ \cite{Neubert:1993um}.
$M_n$ in \eqref{SF} are related to the moments of the photon energy spectrum: $\langle y^k \rangle  = \frac{1}{\Gamma} \, \int_0^{1 - z} \, d y \, y^k \, \frac{d \Gamma}{d y}$.
Using \eqref{dGdy} we have:
\begin{align}
\label{moments}
\langle y^k \rangle = \sum_{n = 0}^\infty \, \frac{M_n}{n!} \, \int_0^{1 - z} \, d y \, y^k \, \delta^{(n)} ( 1 - z - y ) = \sum_{j = 0}^k \, \binom{k}{j} \, ( 1 - z )^{k - j} \,M_j \;,
\end{align}
\vspace{-0.7cm}
\begin{align}
\label{stat}
\langle y \rangle =  ( 1 - z ) \, + M_1  \;,
\hspace{0.5cm}
\langle y^2 \rangle =  ( 1 - z )^2 \,  + 2 \, ( 1 - z ) \, M_1 + M_2 \;,
\hspace{0.5cm}
\sigma_y^2 = \langle y^2 \rangle - \langle y \rangle^2 \;. 
\end{align}
The moments of the measured photon energy spectrum can be used to determine  the HQET parameters as well as the $\cos \theta_P$ distribution \eqref{angular}.

Let us describe  the effect of the Fermi motion.
Each order in the $1 / m_b$ expansion, the photon energy spectrum obtained by the OPE corresponds to a monochromatic  line.
At LO, the line is  at $y = 1 - z$,   the next terms correspond to a  displacement of this position.
The convolution \eqref{convolution} with the shape function provides the smearing of the spectrum.

The shape function is a nonperturbative quantity to be determined by methods such as lattice QCD, or which must be suitably parametrized. 
We adopt a different point of view.
Including infinite terms, the sum in Eq. \eqref{dGdy}  coincides with $S_s(y)$ in \eqref{SF}. 
In the sum, the first term corresponds to a monocromatic line at the zero of the $\delta$ function, with  $\langle y \rangle = 1 - z $ and $\sigma_y^2 = 0$, the LO results in Eqs. \eqref{stat}.
We write $\delta( b - y ) = \lim_{\sigma_y \to 0} \, \frac{1}{\sqrt{2 \, \pi} \,  \sigma_y} \, e^{- \frac{ ( b - y )^2}{2 \,\sigma_y^2}}$, with $b = ( 1 - z ) = \langle y \rangle_\text{LO}$ and $\sigma_y^2$  at each order  in $1 / m_b$,  starting from $1 / m_b^{2}$. 
For $m_b \to \infty$ the condition $\sigma_y \to 0$ reproduces the partonic result. 
Our ansatz consists in substituting
\begin{align}
\label{Ss0}
S_s(y) & = \sum_{n = 0}^{\infty} \, \frac{M_n}{n!} \, \delta^{(n)} ( 1 - z - y )
\hspace{0.5cm}
\to
\hspace{0.5cm}
S_{s }(y) = \sum_{n = 0}^{\infty} \, \frac{M_n}{n!} \,(-1)^n \, \frac{d^n}{dy^n} \, \frac{1}{\sqrt{2 \, \pi} \, \sigma_y} \, e^{- \frac{ ( b - y )^2}{2 \, \sigma_y^2}} \;.
\end{align}
Using the representation of the  Hermite polynomials $H_n(x) = (-1)^n \,  e^{x^2} \, \frac{d^n}{dx^n} \, e^{-x^2}$,
we obtain:
\begin{align}
\label{Ss}
S_{s }(y) = \frac{1}{\sqrt{2 \, \pi} \,\sigma_y} \, \sum_{n = 0}^{\infty} \, \frac{M_n}{n!} \, \left(- \frac{1}{\sqrt{2} \, \sigma_y} \right)^n \, e^{- \frac{ ( b - y )^2}{2 \,\sigma_y^2}} \, H_n \, \left( \frac{ b - y}{\sqrt{2} \,\sigma_y} \right) \;.
\end{align}
Denoting  by $\langle y^k \rangle_{\cal N}$ the moments computed using this expression for the spectral function,
\begin{align}
\label{momN}
\langle y^k \rangle_{\cal N} = \int_0^{y_{\rm max}} \, dy \,  y^k \,  S_{s }(y) \;, \hspace{0.5cm} \text{where} \hspace{0.5cm} \lim_{\sigma_y \to 0} \, \langle y^k \rangle_{\cal N} = \langle y^k \rangle \;,
\end{align}
with  $\langle y^k \rangle$  in \eqref{moments}.
While $b$ is fixed to the LO result for $\langle y \rangle$, $\sigma_y$ can be determined at an arbitrary order in the $1 / m_b$ expansion.
At odds with most models proposed for the shape function, by construction  the ansatz \eqref{Ss} can include all moments $M_n$ once they are computed.

\begin{figure}[!h]
\begin{center}
\sidecaption
\includegraphics[width = 0.5\textwidth]{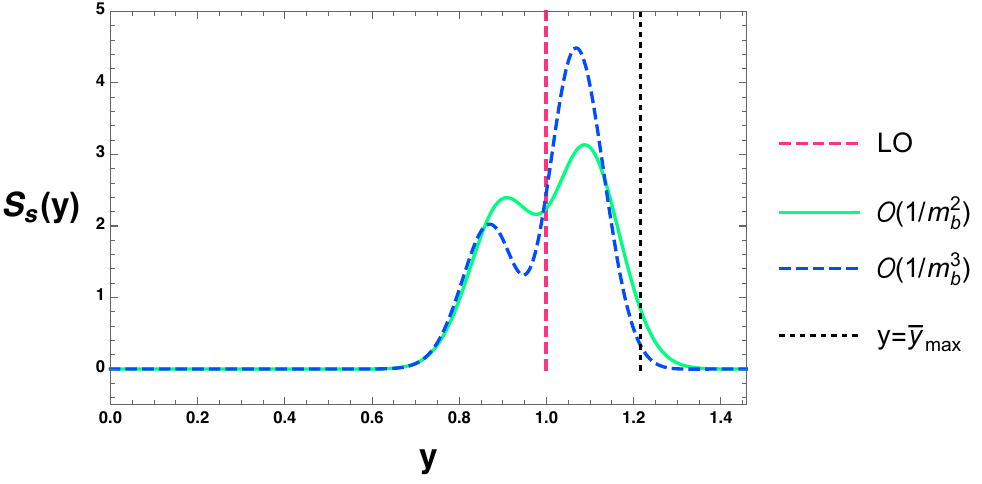}
\includegraphics[width = 0.5\textwidth]{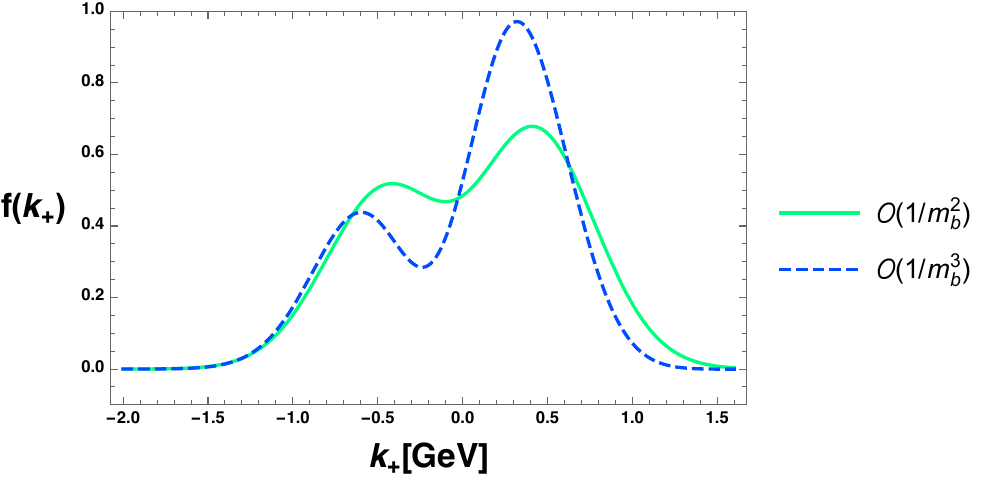}
\vspace{-0.7cm}
\caption{\small Function $S_s(y)$ (left) and  shape function $f(k_+)$ (right) obtained using Eq. \eqref{Ss} up to $n=3$.}
\label{spectrumSF}
\end{center}
\end{figure}

In Fig. \ref{spectrumSF} we plot the spectral function obtained at LO, ${\cal O}(m_b^{- 2})$ and  ${\cal O}(m_b^{- 3})$.
The maximum  photon energy is larger than the partonic result $y = 1 - z$.
In the same figure we plot the shape function obtained from \eqref{intfk} at ${\cal O}(m_b^{- 2})$ and ${\cal O}(m_b^{- 3})$.
As a consequence of broadening the spectrum through the substitution in \eqref{Ss0},  there is a tail  exceeding the physical end-point ${\bar y}_{\rm max} \simeq \frac{m_{\Lambda_b}}{m_b} \, (1-z)$, and  a tail in the shape function exceeding $k_+^{\rm max} = m_{\Lambda_b} - m_b$.
This is a spurious effect of the truncation.
When higher orders in the HQE are included, the area below such tails approaches to zero. 

\vspace{-1cm}

\section{Conclusions}

The HQE has been exploited to compute the inclusive  decay width induced by the $b \to s \, \gamma$ transition for a beauty baryon.
The  differential width in the photon energy and in $\cos \theta_P$ allows to construct new observables  with respect to mesons.
The calculation has been carried out at ${\cal O}(m_b^{- 3})$ for non-vanishing $m_s$,  using the baryon matrix elements  determined in  \cite{Colangelo:2020vhu}.
For  the singular terms appearing in the spectrum we have proposed a treatment that can be systematically improved  including higher order terms in the expansion.

\vspace{-0.2cm}

\section*{Acknowledgments}

I thank P. Colangelo and F. De Fazio for collaboration.
This study has been carried out within the INFN project (Iniziativa Specifica) SPIF.

\nocite{}

\bibliographystyle{woc}
\bibliography{refsFFP6}

\end{document}